\documentclass[a4paper,structabstract]{aa}  
\usepackage{graphicx}
\usepackage{txfonts}

\usepackage{natbib}
\bibpunct{(}{)}{;}{a}{}{,}

\def\depthfin{0.109$\pm$0.025}
\def\depthform{0.1089$\pm$0.01}
\def\depthpb{0.1089$^{+0.024}_{-0.025}$}
\def\TBfin{2136$^{+150}_{-170}$}
\def\sigfin{4}
\def\sigform{11}

\begin{document}

   \title{The GROUSE project II: Detection of the Ks-band secondary eclipse of exoplanet HAT-P-1b\thanks{Photometric time-series are only available in electronic form at the CDS via anonymous ftp to cdsarc.u-strasbg.fr (130.79.128.5) or via http://cdsweb.u-strasbg.fr/cgi-bin/qcat?J/A+A/}}
   \titlerunning{The GROUSE project II: The secondary eclipse of HAT-P-1b}
   \authorrunning{De Mooij et al.}
   \author{E.J.W. de Mooij\inst{1}, R.J. de Kok\inst{2}
           S.V. Nefs\inst{1}
           \and
           I.A.G. Snellen\inst{1}}
   \institute{Leiden Observatory, Leiden University, Postbus 9513, 2300 RA, Leiden, 
               The Netherlands;    \email{demooij@strw.leidenuniv.nl}
             \and SRON Netherlands Institute for Space Research, Sorbonnelaan 2, 3584 CA Utrecht, The Netherlands;    }           
   \date{  }

\abstract
{Only recently it has become possible to measure the thermal emission from hot-Jupiters at near-Infrared wavelengths using ground-based telescopes, by secondary eclipse observations. This allows the planet flux to be probed around the peak of its spectral energy distribution, which is vital for the understanding of its energy budget.}
{The aim of the reported work is to measure the eclipse depth of the planet HAT-P-1b at 2.2$\mu$m. This planet is an interesting case, since the amount of stellar irradiation it receives falls in between that of the two best studied systems (HD209458 and HD189733), and it has been suggested to have a weak thermal inversion layer.}
{We have used the LIRIS instrument on the William Herschel Telescope (WHT) to observe the secondary eclipse of HAT-P-1b in the K$_s$-band, as part of our Ground-based secondary eclipse (GROUSE) project. The observations were done in staring mode, while significantly defocusing the telescope to avoid saturation on the K=8.4 star. With an average cadence of 2.5 seconds, we collected 6520 frames during one night.}
{ The eclipse is detected at the 4-$\sigma$ level, the measured  depth being \depthfin\%. The uncertainties are dominated by residual systematic effects, as estimated from different reduction/analysis procedures. The measured depth corresponds to a brightness temperature of \TBfin K. This brightness temperature is significantly higher than those derived from longer wavelengths, making it difficult to fit all available data points with a plausible atmospheric model. However, it may be that we underestimate the true uncertainties of our measurements, since it is notoriously difficult to assign precise statistical significance to a result when systematic effects are important.}
{}


\maketitle

\section{Introduction}
Measurements of the secondary eclipse of an exoplanet, the moment it passes behind its host star, allow us to probe the properties of the atmosphere on the day-side of the planet. The first successful secondary eclipse measurements have been obtained with the Spitzer Space Telescope ~\citep{charbonneauetal05, demingetal05}, followed by many more secondary eclipse measurements from 3.6$\mu$m to 24$\mu$m ~\citep[e.g.][]{knutsonetal08, machaleketal08}, see also the review by~\cite{deming09iau}.

These Spitzer observations indicate that some planets exhibit a thermal inversion in their atmospheres~\citep[e.g.][]{knutsonetal08}. This inversion is apparent when molecular bands in the infrared change from absorption to emission, resulting in a different shape of the planet's spectral energy distribution. \cite{fortneyetal08} and \cite{burrowsetal08} have proposed that such an inversion layer could be due to the presence of a strong optical absorber high in the planet's atmosphere, depending on the amount of stellar radiation the planet receives. Only for the planets receiving very high levels of irradiation, the hypothetical absorbing compound can stay in the gas-phase high up in the atmosphere, absorbing the stellar light very efficiently and causing the thermal inversion.  As a possible absorber, \cite{hubenyetal03} suggested TiO and VO, although recent work~\citep{spiegeletal09} show that it might be difficult to keep these molecules in the gas phase. At lower levels of stellar irradiation, the compound possibly condenses out and is subsequently removed from the higher layers of the atmosphere. \cite{fortneyetal08} dub these two classes pM and pL respectively, in analogy with L and M stellar dwarfs.

With the increase in the number of exoplanets studied with the Spitzer Space Telescope, it became clear that the above scheme cannot be solely dependent on the level of stellar irradiation.  There are planets apparently exhibiting an inversion layer which receive less light from their host-star than required to keep the proposed absorber in the gas-phase~\citep[e.g. XO-1b, ][]{machaleketal08}, while there are also planets that receive very high levels of irradiation, which do not apparently exhibit an inversion layer~\citep[e.g. TrES-3b, ][]{fressinetal10}. Recently,~\cite{knutsonetal10}, showed that there appears to be a trend between the presence of an inversion layer and the stellar activity (as determined from the calcium lines), with only planets around stars with low activity exhibiting an inversion layer. \cite{madhusudhan} showed that the inference of a thermal inversion is not robust, and can also depend on the chemical composition of the atmosphere.

Recently, several groups have presented measurements for a number of exoplanets in the near-infrared using ground-based telescopes~\citep{demooijandsnellen09, singandlopezmorales09, gillonetal09,rogersetal09, andersonetal10, alonsoetal10, gibsonetal10, crolletal10a, lopezmoralesetal10, crolletal10b, crolletal11}. Data in this wavelength region are very interesting because they probe the thermal emission at, or close to, the peak of the spectral energy distribution of hot-Jupiters, which is vital for the understanding of their energy budgets. 

Here we present the second result from the GROUnd-based Secondary Eclipse project (GROUSE), which aims to use telescopes at a variety of observatories for exoplanet secondary eclipse observations. As part of this survey we have already published our detection of the secondary eclipse of TrES-3b~\citep{demooijandsnellen09}, which we retrospectively include as paper~I.

In this paper we present observations of the secondary eclipse of the hot-Jupiter HAT-P-1b~\citep{bakosetal07} in K$_s$-band as obtained with the Long-slit Intermediate Resolution Infrared Spectrograph~\cite[LIRIS;][]{LIRIS02} instrument on the  William Herschel Telescope (WHT). HAT-P-1b, with a mass of 0.52M$_{jup}$ and a radius of 1.2R$_{jup}$, orbits its G0V stellar host with a period of P=4.5 days at a distance of 0.055 AU. This large orbital separation places HAT-P-1 on the lower edge of the proposed pM/pL transition boundary of~\cite{fortneyetal08}.  Recent observations of this planet by~\cite{todorovetal10}, who used the IRAC instrument on the Spitzer Space Telescope to determine the infrared brightness of HAT-P-1, showed evidence for a weak inversion layer. Its host star appears not to be very active~\citep{knutsonetal10}, which provides an alternative explanation for the planet's inversion layer, in the case that stellar variability is one of the driving factors for the structure of the atmosphere.

In Sec.~\ref{sec:obs_dr} we present our observations and data-reduction. In Sec.~\ref{sec:res} we will present and discuss the results, and in Sec.~\ref{sec:concl} we will give our conclusions.

\section{Observations, data reduction and analysis}\label{sec:obs_dr}

The secondary eclipse of HAT-P-1 was observed on October 2, 2009 using the imaging mode of the Long-slit Intermediate Resolution Infrared Spectrograph~\cite[LIRIS;][]{LIRIS02} instrument on the  William Herschel Telescope (WHT) on La Palma. The expected time of mid-eclipse was 21$^h$:55$^m$UT. The observations started at 19$^h$:59$^m$UT and lasted for 6.5 hours, although $\sim$15 minutes were lost because the observer got locked up in the bathroom due to a broken door handle. Since this caused a gap in the photometry, we excluded the final 630 frames from further analysis. The exposures were taken in sequences of 50 or 100 frames. The first two frames of every sequence are known to suffer from a reset-anomaly\footnote{http://www.ing.iac.es/astronomy/instruments/liris/liris\_cookbook.pdf}, visible by a very strong gradient in the background. We excluded the frames affected by the anomaly from further analysis.

The LIRIS detector was windowed to one sub-array of 512 by 512 pixels in order to increase the observing cadence. The pixel-scale of LIRIS is 0.25 arcsec per pixel, hence the field of view for these windowed observations was 128 by 128 arcseconds. Since HAT-P-1b has a slightly brighter companion star nearby, the field of view was enough to allow us to observe both HAT-P-1 (K=8.86) and the reference star, ADS 16402 A (K=8.41), simultaneously.

The observations were performed in staring mode, keeping the stars as much as possible at the same position on the chip. Despite guiding, a slow drift of 3 pixels over 4 hours is visible, and subsequently a sudden jump of 5 pixels in right ascension occurs. The stars also slowly drifted by 2 pixels in declination during the night.

The exposure times were varied between 1.3 and 2 seconds, in order to keep the flux-levels of the star in the linear regime of the detector. To allow these relatively long exposure times for these rather bright stars, we strongly defocused the telescope. This resulted in a donut-shaped PSF with a diameter of 15 pixels (3.75 arcsec). Despite the defocus, the PSF was still small enough to avoid overlap between the target and the reference star (which are separated by $\sim$11.2 arcsec), although the diffraction spikes originating from the support of the secondary mirror did overlap for certain periods (see section~\ref{sec:spokes}).

\subsection{Crosstalk, non-linearity corrections and flat-fielding}\label{sec:nlct}
\begin{figure}
\centering
\includegraphics[width=8cm]{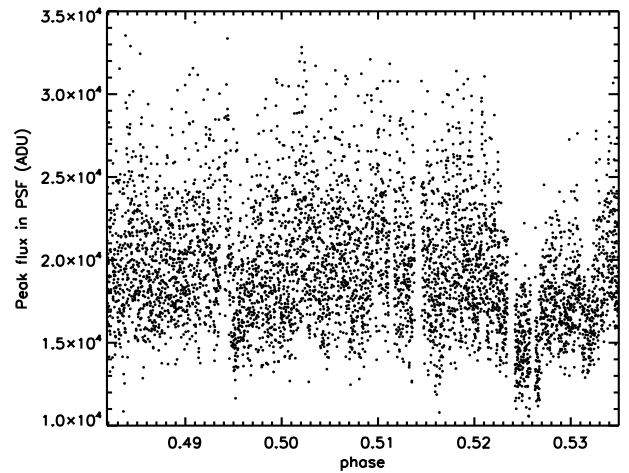}
\caption{The flux in the highest pixel of the PSF of the brightest star as a function of time.}
\label{fig:peakflux}
\end{figure}
As a first step in the data reduction we removed the intra-quadrant crosstalk, which is present as an additional flux in each pixel at the level of 10$^{-5}$ of the total flux along its row. We subsequently performed a non-linearity correction, for which we obtained a new set of measurements in August 2010. A plot of the value of the highest pixel in the PSF of the brightest star is shown in Fig.~\ref{fig:peakflux}. For the non-linearity correction we observed several sequences of K$_s$-band dome-flats with varying exposure times to which we subsequently fitted a second order polynomial to the non-linearity for each pixel of the detector. Our current results show that the detector is non-linear for all fluxes, reaching $\sim$8.5\% at 30000 ADU, above this flux level the non-linearity increases rapidly. We used this non-linearity solution to correct all our frames. We excluded all frames with the peak flux of the brightest star above 30000 ADU.

All the frames were flatfielded using a flatfield created from a set of dome flats, both with the dome-lights on and off to remove both the effects of emission from the telescope structure, as well as structure due to the dark-current from the detector. The flat-field images were corrected for the intra-quadrant crosstalk and non-linearity in the same way as the science images.

\subsection{Removal of bad-pixels}
During most of the observations, one hot-pixel is located within the PSF of the target. This hot-pixel was corrected for by replacing it with the flux of the corresponding pixel in the reference star, scaled by the flux ratio of the stars. Since the position difference between the two stars is not an integer, and the precise flux in inner edge of the donut is a strong function of position, we interpolated the PSF of the reference star to the same grid as HAT-P-1. before measuring the flux in the corresponding pixel.

\subsection{Background subtraction}
\begin{figure}
\centering
\includegraphics[width=8cm]{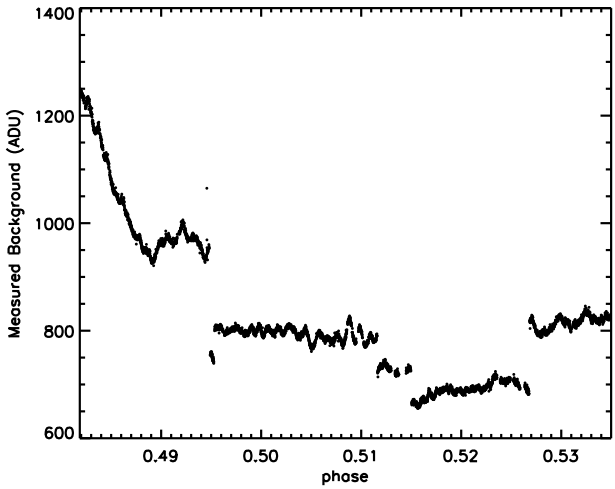}
\caption{The variation of the sky background during the night.}
\label{fig:bgflux}
\end{figure}
The background level during the observations varied, and the variation of the background per pixel is shown in Fig.~\ref{fig:bgflux}. A background map was constructed from a separate set of observations of a different field taken on the same night. After subtracting this background map from the images,  a small gradient was still apparent along the y-axis, which we believe is due to different exposure times (and dark currents) for the images used for the creation of the background map and the science images. To remove this gradient, we first masked out all the stars and bad-pixels and subsequently determined the mean along a row of the detector, rejecting outliers deviating by more than 3 times the standard deviation from the median of the row. The resultant profile along the y-axis was then fitted with a polynomial, in order to create a smooth background map, which was subsequently subtracted from all the columns of the image.  

The overall impact of the subtraction of a background image on the results is small, but not subtracting the background image results in an increased noise-level.

\subsection{Diffraction spokes from the secondary mirror support}\label{sec:spokes}
\begin{figure}
\centering
\includegraphics[width=8cm]{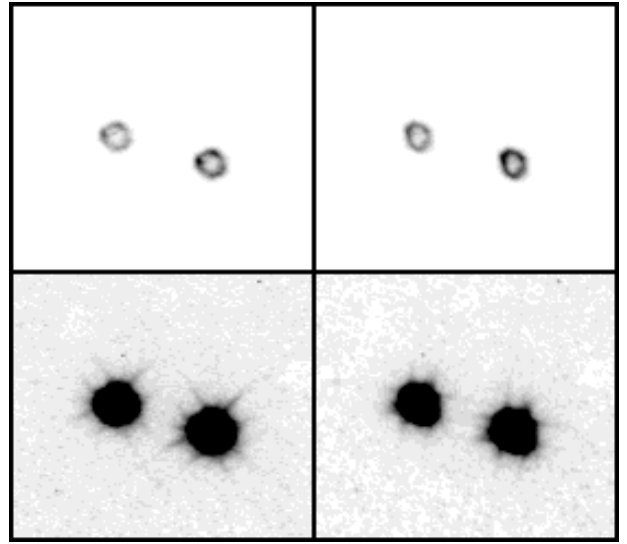}
\caption{Sample of PSFs for our observations at high and low count-levels for different mount rotator angles. Left panels: mount rotator angle of -79.2 degrees. Right panels: +55.0 degrees. The spokes caused by the support structure of the secondary mirror can be clearly seen to rotate due to the alt-az mounting of the telescope.}
\label{fig:spokes}
\end{figure}
The close proximity of the reference star causes the diffraction pattern from the support structure for the secondary mirror to overlap with the other star's PSF. Since the WHT is on an alt-az mount, the instrument is rotated to maintain the same position angle on the sky. This causes the diffraction spikes at certain times to overlap, depending on the position-angle of the mount. In Fig.~\ref{fig:spokes} two images are shown, on the left side with a rotator angle of -79.2 degrees, with the diffraction spikes not overlaping, while in the right side the rotator angle was +55 degrees, where the diffraction spikes do overlap. To account for this,  the amount of flux in a region offset by the same amount as the separation between the two stars was measured, and subsequently subtracted from the measured flux values of the stars, after accounting for the difference in flux levels.

\subsection{Aperture Photometry}
Subsequently, we performed aperture photometry on the two stars, with a radius of  14 pixels (3.5 arcsec), chosen to maximise the flux, but minimise the influence of the background. Any residual background was determined by measuring the mean flux in two boxes located 60 pixels (15 arcsec) above and below the midpoint between the two stars, with a  box-size of is 131 by 41 pixels (32.75 by 10.25 arcsec). In these boxes, we excluded bad-pixels from the background-determination, as well as masked the areas that can be affected by the diffraction spikes from the secondary mirror support.

\subsection{Correction for systematic effects}
\begin{figure}
\centering
\includegraphics[width=8cm]{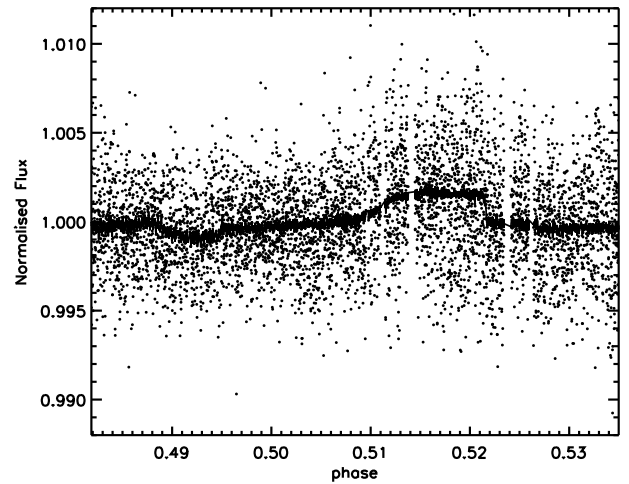}
\caption{Data points show the uncorrected lightcurve of HAT-P-1b with the WHT on La Palma. The solid line indicates the best fitting model where we fit simultaneous for both the systematic effects as well as the eclipse depth. }
\label{fig:orig_sececl}
\end{figure}
\begin{figure}
\centering
\includegraphics[width=8cm]{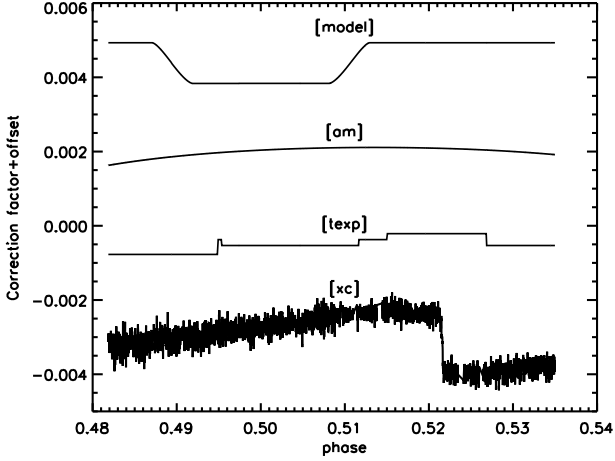}
\caption{Different components fitted to the lightcurve due to the eclipse and systematic effects. Shown are, from top to bottom the eclipse model, the airmass, the exposure time and the x position on the detector. The different parameters are offset for clarity.}.
\label{fig:sec_ecl_param}
\end{figure}
\begin{table}
\caption{Parameters for HAT-P-1. References: (1)~\cite{winnetal07}; (2)~\cite{johnsonetal08}; (3) this work;  (4)~\cite{todorovetal10}}
\label{tab:sysparam}
\centering

\begin{tabular}{c c c}
\hline\hline
 parameter & Value &  Reference \\
\hline 
Semi-major axis  a (au)     & 0.0551$\pm$0.0015     & (1) \\
R$_*$ (R$_\odot$)            & 1.115$\pm$0.043       & (1) \\
R$_p$ (R$_{jup}$)            & 1.204$\pm$0.051       & (1) \\
i (deg)                     & 86.22$\pm$0.24        & (1) \\
P (days)                    & 4.4652934$\pm$0.00072 & (2) \\
\hline
b                           & 0.701$\pm$0.023       & (1) \\
R$_p$/R$_*$                 & 0.11295$\pm$0.00073   & (2) \\
\hline
Eclipse depth at 2.2$\mu$m  & \depthfin\%           & (3) \\
Eclipse depth at 3.6$\mu$m  & 0.080$\pm$0.008\%     & (4) \\
Eclipse depth at 4.5$\mu$m  & 0.135$\pm$0.022\%     & (4) \\
Eclipse depth at 5.8$\mu$m  & 0.203$\pm$0.031\%     & (4) \\
Eclipse depth at 8.0$\mu$m  & 0.238$\pm$0.040\%     & (4) \\
\hline
\end{tabular}
\end{table}
The lightcurve of HAT-P-1b, corrected for the hot-pixel, diffraction spikes, and normalised with that of the reference star, is shown in Fig.~\ref{fig:orig_sececl}.  
Although the eclipse is already visible, the lightcurve is still clearly influenced by systematic effects, which we found to be correlated with the x-position  on the detector, the exposure time and the geometric airmass. We fitted for these effects simultaneously with the eclipse depth. Other parameters, such as the background level, were tried, but resulted in no difference in the final lightcurve. For the model of the secondary eclipse, we used the formalism from~\cite{mandelandagol02}, where all the orbital parameters were kept fixed to the literature values (given in Table~\ref{tab:sysparam}), with the limbdarkening set to 0, and the timing of the secondary eclipse fixed. We did not fit for a time offset, because \cite{todorovetal10} already showed that the timing of the secondary eclipses measured using the Spitzer Space Telescope shows no indication of any offset, using higher signal-to-noise data.  

We performed the fit both using the standard multi-linear regression algorithm in IDL (regress.pro) as well as using a Monte-Carlo Markov Chain (MCMC) analysis.
After an initial fit, we performed a clipping of the residuals by excluding all points within bins of 38 pixels, for which the binned value deviate by more that 1.5$\cdot$10$^{-3}$. In this way we removed 228 unbinned points. After this clipping we performed an MCMC analysis.
For the MCMC analysis, we created 5 chains with a length of 2 million steps each, using different starting positions. We trimmed away the first 200.000 points of each chain, to make sure that the initial starting point no longer influence the measured parameters. 

The best fitting model, overplotted on the data, is shown in Fig.~\ref{fig:orig_sececl}, while the contributions for the individual components are shown in Fig.~\ref{fig:sec_ecl_param}. 

\section{Results}\label{sec:res}

\begin{figure}
\centering
\includegraphics[width=8cm]{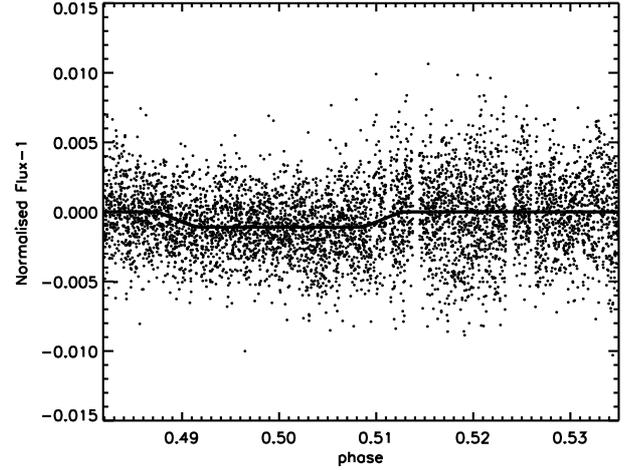}
\caption{The lightcurve for the secondary eclipse of HAT-P-1b corrected for systematic effects. Overplotted is our best fitting eclipse model.}
\label{fig:cor_sececl}
\end{figure}
\begin{figure}
\centering
\includegraphics[width=8cm]{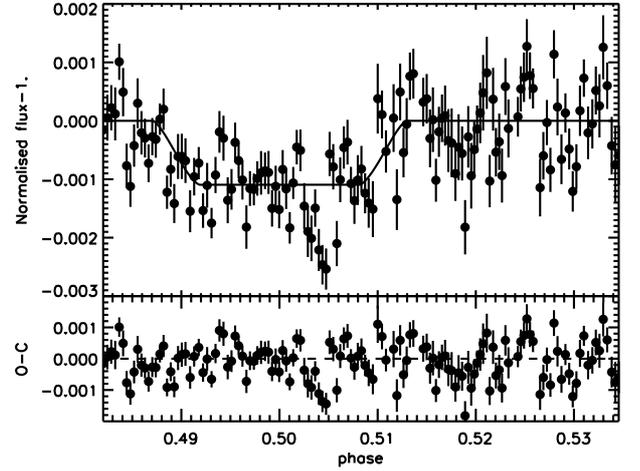}
\caption{The same as Fig.~\ref{fig:cor_sececl}, but binned by 41 points.}
\label{fig:cor_sececl_bin}
\end{figure}

The final, corrected lightcurve is shown in Fig.~\ref{fig:cor_sececl}, with its binned (by 41 points) version in Fig.~\ref{fig:cor_sececl_bin}. Assuming pure, white, Gaussian noise, we find an eclipse depth of \depthform~\% (\sigform-$\sigma$). However, clear residual systematic effects are visible in the binned lightcurve. The level of this red noise was estimated by comparing the standard deviation in the unbinned data with that of gradually increased binned lightcurves. From this we estimate a red noise component at the level of 0.023\%, which we added in quadrature to our eclipse depth uncertainty to become \depthfin~\%. We also performed the residual permuation (``Prayer Bead'') method~\citep[e.g.][]{gillonetal2007}, where we add the best fit model to the residuals, after shifting the residuals by $N$-points. Points are wrapped around when performing the shift. By refitting the data for all possible shifts between 0 and the number of data points, and measuring the best-fit eclipse depth, we can get an estimate for the true uncertainties in the data. We find a depth of \depthpb~\%, which is fully consistent with the uncertainties estimated from the red-noise.

This results overall in an eclipse depth of \depthfin~\%, which corresponds to a brightness temperature in the K$_s$-band of T$_b$=\TBfin K. This brightness temperature is significantly higher than the expected equilibrium temperature of this planet, which lies between T$_{eq}$=1540 K for an albedo of 0 and inefficient energy redistribution from the dayside to the nightside (P$_n$=0), and  T$_{eq}$=1190 K for an albedo of 0.3 and efficient redistribution of energy from the dayside to the nightside (P$_n$=0.5). The higher brightness temperature could be due to the fact that in the K$_s$-band we are looking deeper into the atmosphere, where the layers are warmer.

\cite{todorovetal10} observed the secondary eclipse of HAT-P-1b with the IRAC instrument on board the Spitzer Space Telescope, at 3.6, 4.5, 5.8 and 8.0 $\mu$m, and find eclipse depths of 0.080$\pm$0.008\%, 0.135$\pm$0.022\%, 0.203$\pm$0.031\% and 0.238$\pm$0.040\% in the different bands respectively. 
Combining our own measurement of the secondary eclipse in K$_s$-band with the IRAC measurements, we can construct a spectral energy distribution (SED) from the near-infrared to the mid-infrared, which is shown in Fig.~\ref{fig:models}, and compare it to atmospheric models.

\subsection{Atmospheric models}
\begin{figure}
\centering
\includegraphics[width=8cm]{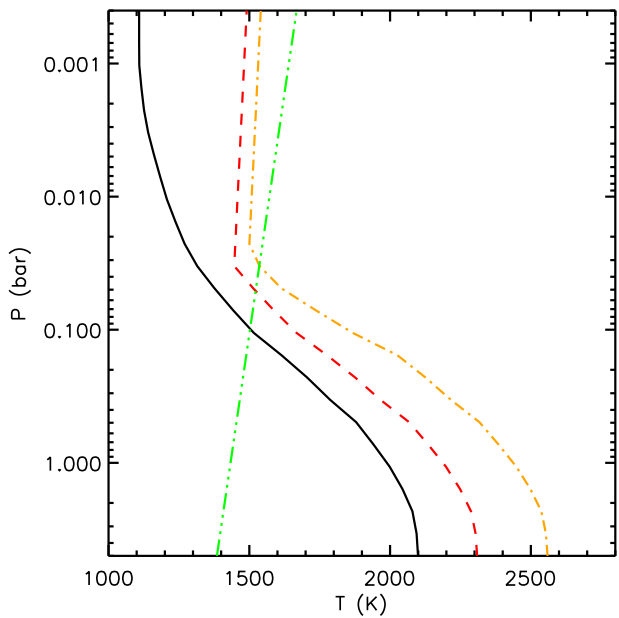}
\caption{The temperature-pressure (T-P) profile for the our atmospheric models used for modelling the secondary eclipse depth. The black (solid) line shows the uninverted model from~\cite{fortneyetal08}, the green (dashed-triple dotted) line is for a model with an ad-hoc temperature inversion as in~\cite{todorovetal10}, the red (dashed) line shows the temperature pressure profile for an atmosphere with a normal troposphere and an inversion, and the orange (dash-dotted) line is for an atmosphere with Venus-like clouds. The temperatures for the latter profiles are chosen to provide a reasonable fit to the IRAC bands longward of 4$\mu$m}
\label{fig:tp}
\end{figure}
\begin{figure*}
\centering
\includegraphics[width=16cm]{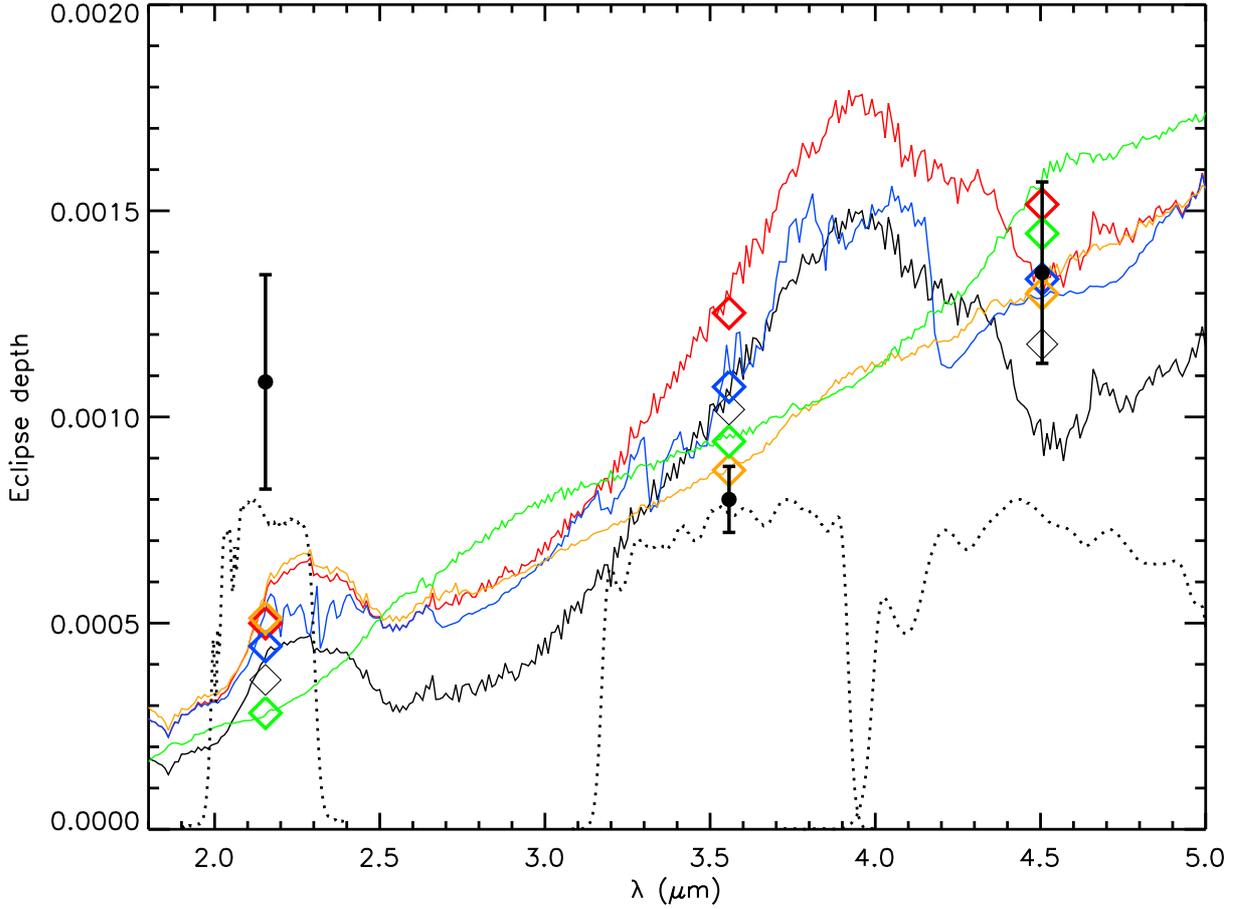}
\caption{Models for the emission from the atmosphere of HAT-P-1b, for the T-P profiles of Fig.~\ref{fig:tp}, using the same colour scheme. The diamonds indicate the expected flux within the observed bands for the different models. The points with errorbars indicate the observational data. The dotted lines indicate the transmission curves of the different bands. In addition, a blue (long dashed) curve is shown for an atmosphere with the same T-P profile as the red (short dashed) curve, but with a high concentration of methane and carbon-dioxide (1$\cdot$10$^{-4}$ and 2$\cdot$10$^{-5}$ respectively).}
\label{fig:models}
\end{figure*}
\begin{figure*}
\centering
\includegraphics[width=16cm]{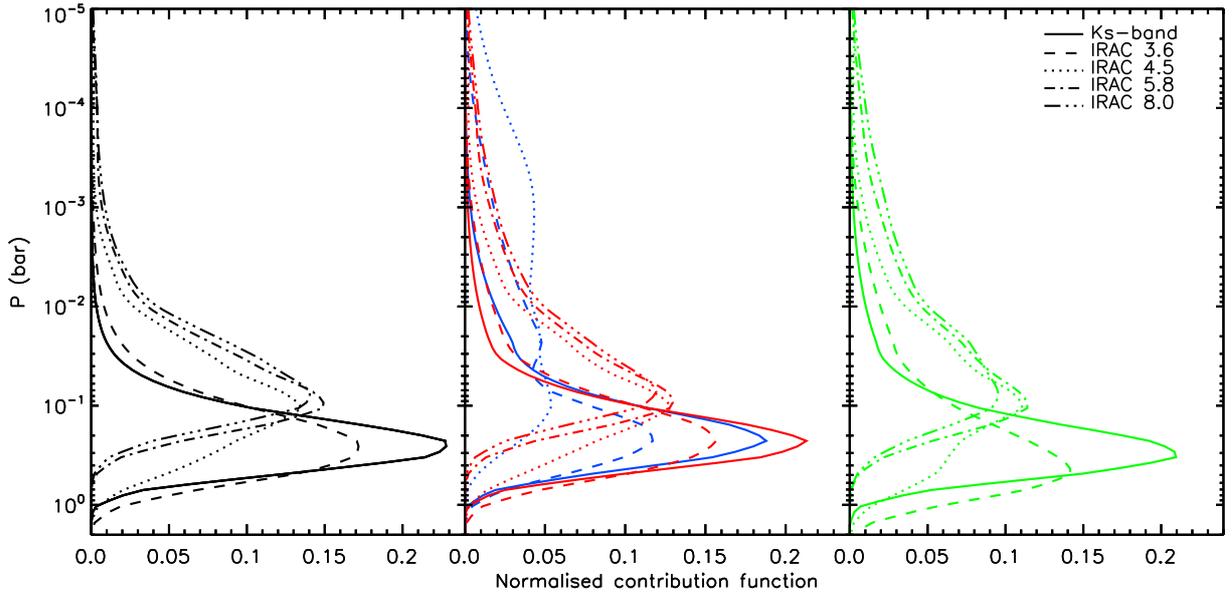}
\caption{Normalised contribution functions for the atmospheric emission of HAT-P-1b in the K$_s$-band and the different IRAC channels, using the same colour scheme as in Figs.~\ref{fig:tp}~and~\ref{fig:models}, although the different linestyles are now used for different bands. The distribution functions are normalised by their sum. The left panel shows the contribution functions for the non-inverted model from~\cite{fortneyetal08}, the middle panel shows the contribution functions for the models with both a troposphere and an inversion layer, and the right panel shows the same for the ad-hoc temperature inversion used in~\cite{todorovetal10} }
\label{fig:contrib}
\end{figure*}

We modelled the thermal emission spectra for a few ad hoc cases to try to understand what type of atmosphere would be needed to fit the measured eclipse depths, including our new measurement in Ks-band. We used a radiative transfer model that can calculate multiple scattering of the thermal radiation, since we included a case with scattering clouds (see below). The scattering is calculated using the doubling-adding method, which makes use of the fact that one can calculate the reflection and transmission properties of a combination of two atmospheric layers from the properties of the individual layers. For each layer in the model atmosphere we started with an optically thin layer with the same scattering properties as the model layer. For the optical thin case analytical expressions are available to calculate the layer transmission and reflection. The thin layer was then doubled several times to match the real optical depth of the model layer. Subsequently, the different model layers were added to calculate the spectra for the entire inhomogeneous atmosphere. For more explanations we refer to~\cite{waubenetal94}, who describe the model in great detail. Spectra were calculated at 10 different emission angles, from which a disc-averaged spectrum was calculated.

We included absorption of water and CO, whose absorption properties were taken from the (old) HITEMP database~\citep{rothmanetal95}.  Although the water data in this database is not the most up-to-date, a line-by-line comparison in K-band with the new HITEMP database~\citep{rothmanetal10} yielded only differences many times smaller than the errors on the measured eclipse depths, which we neglect for this qualitative assessment. For speedy calculations, we pre-computed absorption properties at a range of temperatures and pressure using the correlated-k method~\citep{lacisandoinas91} with a spectral resolution of 0.01 $\mu$m between 1.8 $\mu$m and 5 $\mu$m and 0.1 $\mu$m between 5 $\mu$m and 10$\mu$m, and a Voigt line shape. Assumed mixing ratios of water and CO were 2e-4 and 5e-4 respectively for all cases.

The models extend from 1.8$\mu$m to 10$\mu$m, and were made for different temperature-pressure (T-P) profiles and compositions.  We have made models for the T-P profile for this planet from~\cite{fortneyetal08}, which does not include an inversion layer, as well as for models with an inversion layer. We also varied the composition of the atmosphere and made a model with Venus-like clouds. In Fig.~\ref{fig:tp} we show the temperature-pressure profiles for the different models, and in Fig.~\ref{fig:models} we show the expected eclipse depths within the observed bands for the different models, binned to a resolution of 0.1$\mu$m.

Fig.~\ref{fig:models} shows that an atmosphere calculated from first principles~\citep{fortneyetal08} cannot reproduce the Spitzer measurements, and that a weak inversion is needed~\citep{todorovetal10}. However, a weak inversion throughout the atmosphere produces a very low planet signal in Ks-band, unlike what we derive from our measurements. The reason for this is that Ks-band is a region of very low absorption and hence we see deep in the planet atmosphere, where temperatures are relatively low in the inversion-only case (green lines). Contribution functions for the different wavelengths are shown in Fig.~\ref{fig:contrib}. To increase the eclipse depth in Ks-band the lower atmosphere needs to be made hotter. However, this also increases the eclipse depth in L-band, another spectral region of low absorption (red cureve in Fig.~\ref{fig:models}). L-band emission can be suppressed somewhat by including large amounts of CO2 and CH4 (calculated here from HITEMP and HITRAN 2008~\citep{rothmanetal09} data respectively), but this also reduces the signal in Ks-band (blue curve in Fig.~\ref{fig:models}). Reducing the water or CO abundances also do not produce a better fit. Hence, we cannot fit all data points well using any clear atmosphere with gases that are predicted to be abundant on hot exoplanets. One remedy to fit both the Spitzer data and our Ks-band eclipse depth is to include a Venus-like cloud. The clouds on Venus are made of concentrated sulphuric acid droplets, which have the property that they are strongly scattering below wavelengths of $\sim$3$\mu$m and strongly absorbing above 3 $\mu$m, with only little variation of extinction with wavelength~\citep[e.g.][]{grinspoonetal93}. Because of the scattering nature of the clouds, radiation from Venus' hot lower atmosphere still reaches space at the night side below 3$\mu$m, despite an optically thick cloud layer surrounding the planet. Above 3 $\mu$m the night side emission originates from the clouds. To show this potential effect for HAT-P-1b, we inserted a cloud layer with an optical depth of 1.5 at the tropopause with properties identical to those of Venus' 1-$\mu$m sized cloud-particles. Indeed, brightness temperatures at Ks-band are higher than anywhere else in Fig.~\ref{fig:models} and all data points could potentially be fitted well if the lower atmosphere is made even hotter. However, we could not find a physically plausible candidate for cloud materials that could mimic Venus' clouds on a hot exoplanet. So, at present we do not find any suitable atmosphere that could explain the high Ks-band eclipse depth.

\section{Conclusion}\label{sec:concl}
Using the LIRIS infrared camera on the WHT, we determine the eclipse depth of the extrasolar planet HAT-P-1b in K$_s$-band to be \depthfin\% ($\sim$\sigfin$\sigma$), with the uncertainties in the eclipse depth dominated by residual systematic effects, as estimated from different reduction/analysis procedures. The measured depth corresponds to a brightness temperature of \TBfin K. This brightness temperature is significantly higher than those derived from longer wavelengths, making it difficult to fit all available data points with a plausible atmospheric model. It may be that we underestimate the true uncertainties of our measurements, since it is notoriously difficult to assign precise statistical significance to a result when systematic effects are important.

\begin{acknowledgements}
We are grateful to the staff WHT telescope for there assistance with these observations. The William Herschell Telescope is operated on the island of La Palma by the Isaac Newton Group in the Spanish Observatorio del Roque de los Muchachos of the Instituto de Astrof\'isica de Canarias. 
\end{acknowledgements}
\bibliographystyle{aa} 
\bibliography{demooij_hatp1} 

\end{document}